\documentclass[twocolumn]{aastex631}

\def\msun{{\rm\,M_\odot}}
\usepackage{amsmath} 
\usepackage{multirow}

\begin{document}
\title{How Fast Could Supermassive Black Holes Grow At the Epoch of Reionization?}

\author{Ziyong Wu}
\affiliation{Institute of Astronomy, School of Physics, Zhejiang University, Hangzhou 310037, China}
\affiliation{Purple Mountain Observatory, Chinese Academy of Sciences, 10 Yuanhua Road, Nanjing 210033, China}
\affiliation{School of Astronomy and Space Sciences, University of Science and Technology of China, Hefei 230026, China}

\author{Renyue Cen}
\altaffiliation{renyuecen@zju.edu.cn}
\affiliation{Center for Cosmology and Computational Astrophysics, Institute for Advanced Study in Physics, Zhejiang University, Hangzhou, China}
\affiliation{Institute of Astronomy, School of Physics, Zhejiang University, Hangzhou 310037, China}

\author{Romain Teyssier}
\affiliation{Department of Astrophysical Sciences, Princeton University, Peyton Hall, Princeton, NJ 08544, USA}





\begin{abstract}
Utilizing cosmological hydrodynamic simulations we show that
there is a brief super-Eddington accretion phase in typical 
halos at high redshift, impervious to AGN self-regulation.
However, once having attained a black hole mass of $10^4-10^5\msun$, 
AGN feedback process can self-regulate to guide 
the SMBHs to grow at a significantly slower, sub-Eddington rate. 
By redshift $z\sim 10$ the black hole mass with an initial super-Eddington jump-start is caught up by that in the case with a steady Eddington limited case. 
Thus a continuous Eddington limit case 
represents the fastest possible route to maximally grow SMBHs.
To account for the observed $z=7-10$ quasars with supermassive black holes 
of billions of solar masses, 
our analysis establishes firmer ground for the need of seed masses of 
$10^4-10^5\msun$ that are not grown via an earlier super-Eddington phase. 
\end{abstract}



\section{Introduction} \label{sec:intro}

The formation of supermassive black holes (SMBHs) at high redshift is a fundamental problem in astrophysics \citep[][]{2020Inayoshi}.
The existence of dozens of bright quasars with luminosities $\ge 10^{47}~{\rm erg~s^{-1}}$
at redshift $z>6-7$ \citep[e.g.,][]{2011Mortlock, 2015Wu,2018Banados,2020Yang,2020ARA&A..58...27I,2023ARA&A..61..373F,2024A&A...691A.145M,2025ApJ...980L..29A,2024Natur.628...57F,2025arXiv250408039L,2025arXiv250316596N,2023ApJ...957L...7K,2024arXiv241204983T,2023ApJ...953L..29L,2023ApJ...955L..24G,2023A&A...677A..88B,2024ApJ...965L..21K,2025A&A...693A..50N},
indicating that supermassive black holes (SMBHs) have grown to masses
greater than a few billion solar masses within less than 1~Gyr,
poses one of the greatest challenges to galaxy formation.
Stellar mass black hole seeds cannot grow to the observed masses,
without almost continuous, significantly super-Eddington accretion \citep{2012NatCo...3.1304G,2016MNRAS.458.3047P,2020ARA&A..58...27I}.


Observations of the local Universe show a steep drop
in the abundance of massive black holes below $10^{4-5}\msun$ \cite[e.g.,][]{2007Greene, 2009Goulding}, suggesting a seed black hole mass close to $10^{4-5}\msun$.
If the growth of SMBHs is accompanied by that of galactic stellar bulges,
as evidenced by the correlation between SMBH mass and bulge mass/velocity dispersion
\citep[e.g.,][]{1998Richstone,1998Magorrian,2000Ferrarese,2002Tremaine},
then SMBHs of masses $10^5-10^6\msun$ in many bulgeless massive galaxies \citep[][]{2013Kormendy, 2017She}
may reflect this initial black hole mass with minimal additional growth.
Thus, these two pieces of evidence from local observations provide a tantalizing hint that SMBH has an initial seed mass around $10^4-10^5\msun$.

In this {\it Letter}, we perform detailed simulations and their analysis, varying feedback parameters for both supernovae and SMBHs.
We discover a new, important phenomenon: 
while an early and brief super-Eddington accretion phase in halos of mass $10^9-10^{10}\msun$, when allowed, the black hole grows
at a rate much faster than the Eddington rate, largely independent of the AGN‑feedback strength prescription.
However, once a BH mass of $10^{4-5}\msun$ is reached,
this super-Eddington jump-start is more than compensated for
by a subsequent, slower, sub-Eddington growth rate,
due to excessive earlier feedback during the super-Eddington phase
that rendered a large reduction of the amount of cold gas surrounding the central BH at later times.
By $z\sim 10$, the mass of the SMBH in the more steady,
Eddington-limited case exceeds that of the initially super-Eddington case. Thus, the fastest route to grow SMBH to $z=7-10$ is via the Eddington-limited case and to account for the observed $z=7-10$ quasars, seed masses of $10^4 - 10^5\msun$, not grown via an earlier super-Eddington phase, are required.

\section{METHODOLOGY AND
SIMULATION DETAILS} \label{sec:method}

\begin{figure*}[htbp]
    \center
    \includegraphics[width=0.85\textwidth]{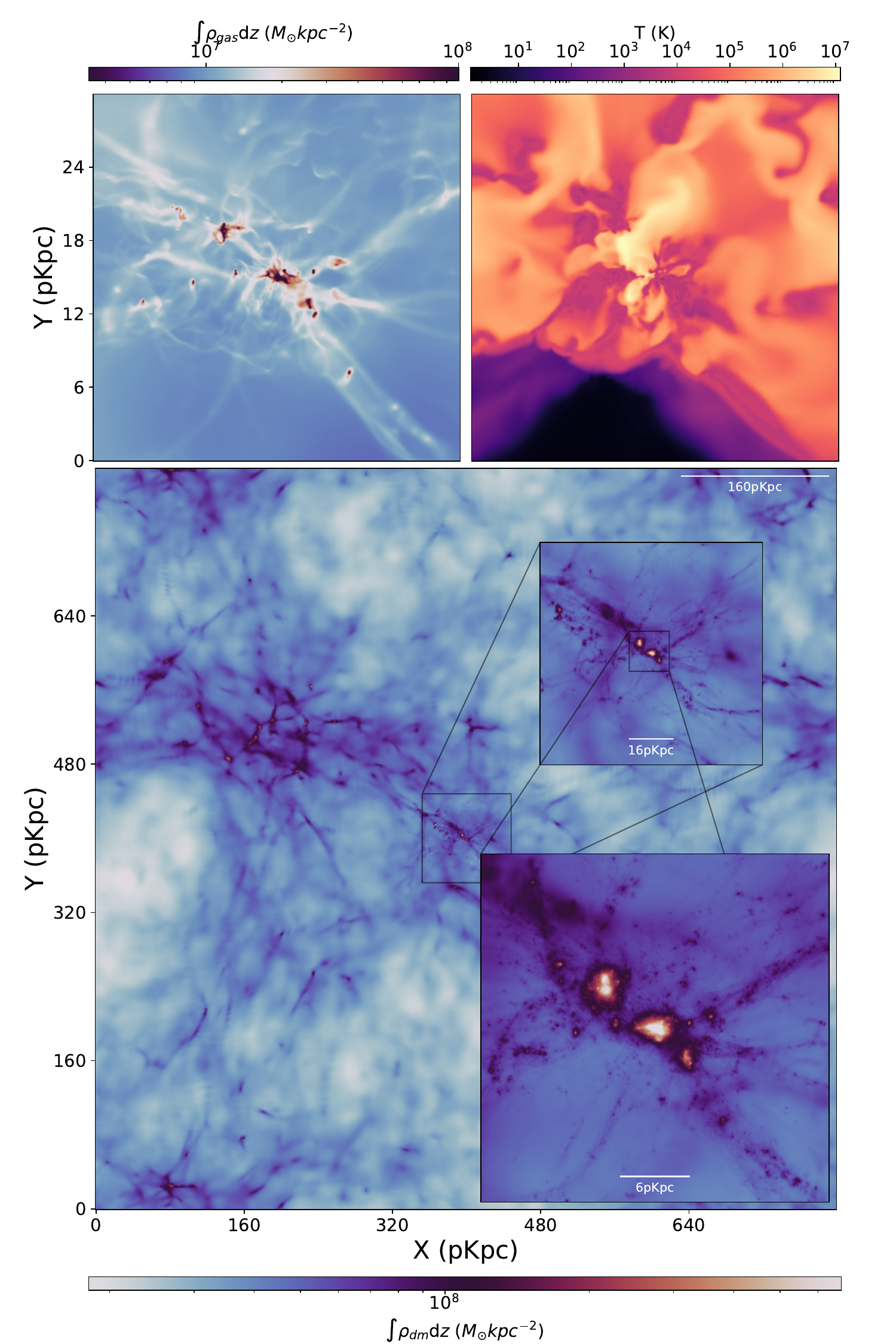} \\
    \caption{The dark matter projected density, the gas projected density as well as the gas temperature plot shown on the bottom panel, upper right panel and upper right panel respectively at $z=10$. All the length unit in the figure is in physical unit. The zoom-in region contains two halos with $2 \times 10^9 \msun < M_{\rm halo} < 5 \times 10^9 \msun$ at $z \sim 10$. }
    \label{fig:simulation_figure}
\end{figure*}

\begin{figure*}[htbp]
    \center
    \includegraphics[width=\columnwidth]{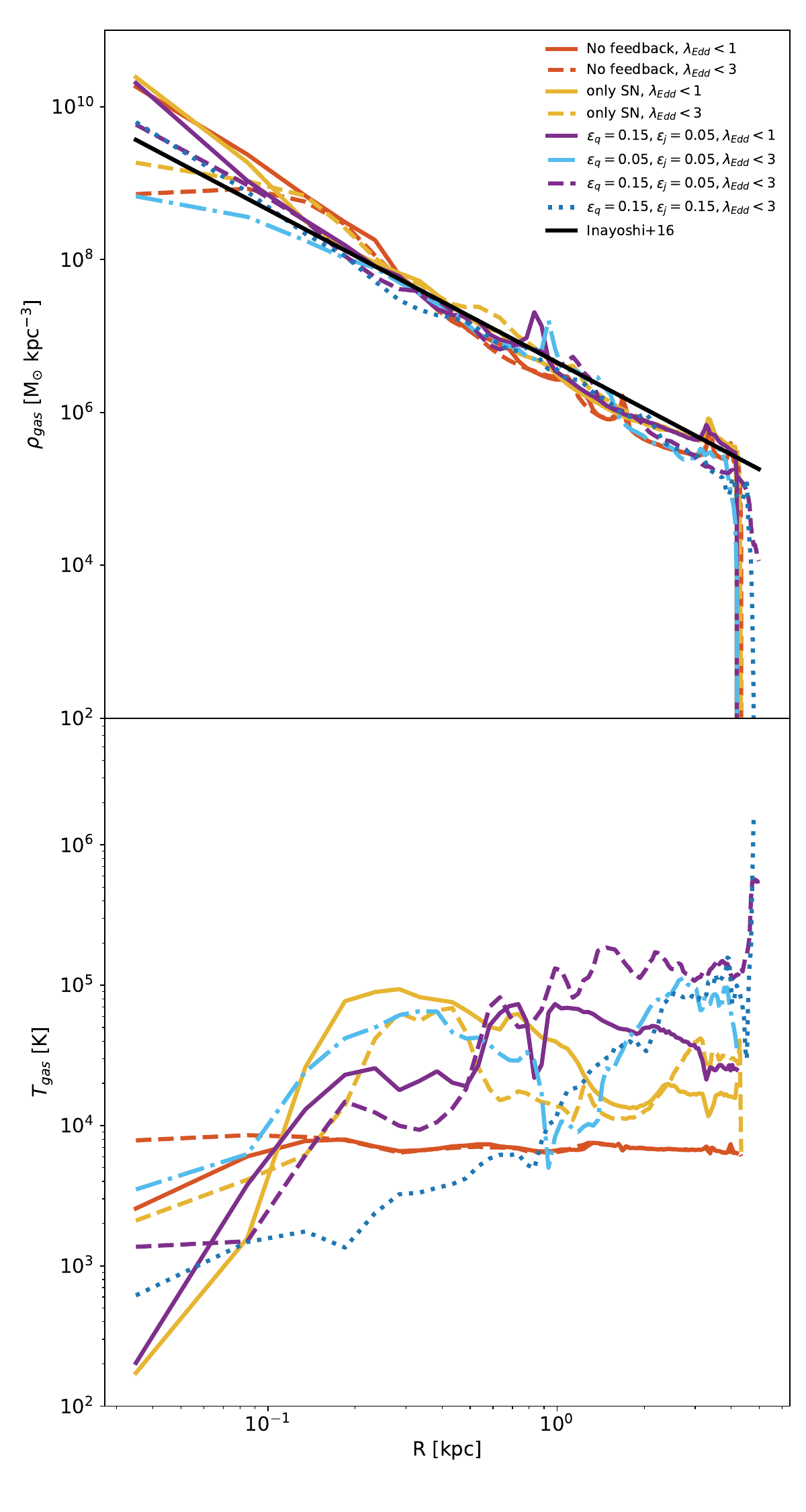} \includegraphics[width=\columnwidth]{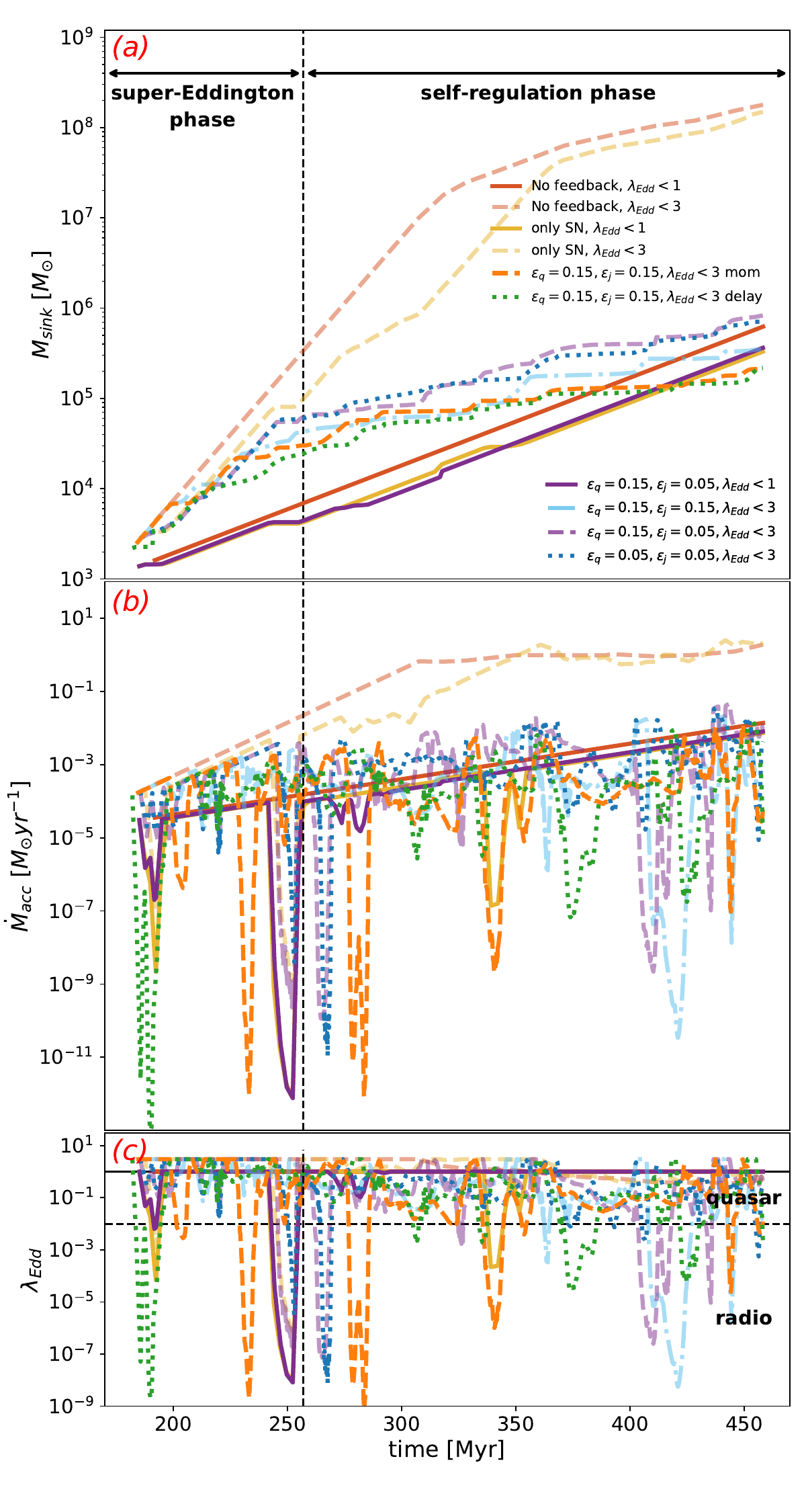}\\
    \caption{{\bf Left:} The gas density and temperature profiles of our zoom-in halo at $z=10$ are shown in the upper and lower panels, respectively. While the gas density profiles are broadly consistent across different simulations, which agree well with the analytical form presented in \citet{2016MNRAS.459.3738I}, the temperature profiles exhibit significant variation. This contrast highlights that, although gravitational collapse governs the overall gas distribution, the thermal state of the gas is highly sensitive to the feedback prescriptions and physical processes included in each simulation. {\bf Right:} (a) The black hole mass growth curves over time are shown, revealing two distinct phases. Once the black hole reaches a mass of $10^4 - 10^5 \msun$ during the super-Eddington accretion phase, AGN feedback self-regulates its growth, causing it to transition to a slower, approximately Eddington-limited rate. We refer to this stage as the self-regulation phase. (b) The evolution of the black hole accretion rate ($\dot{M}_{acc}$, BHAR) over cosmic time is shown, along with (c) the Eddington ratio $\lambda_{Edd}$, and $\lambda_{Edd} = \dot{M}_{acc}/\dot{M}_{Edd}$, which represents the accretion rate relative to the Eddington limit. During the super-Eddington phase, the black hole accretion rate $\dot{M}_{acc}$ remains close to its maximum $3\times \dot{M}_{Edd}$ in the super-Eddington case and $\dot{M}_{Edd}$ in the Eddington-limited case. However, upon entering the self-regulation phase, only the simulations without feedback or with SN-only feedback in the super-Eddington case can sustain the maximum accretion rate. In contrast, in simulations with AGN feedback, the accretion rate inevitably declines to the Eddington limit, for the range of the implemented AGN feedback strength or the suite of SN feedback models employed.}
    \label{fig:simulation_profile}
\end{figure*}

We perform a cosmological hydrodynamic simulation using the adaptive mesh refinement (AMR) code, RAMSES \citep{2002A&A...385..337T}. The initial conditions are generated with the MUSIC software \citep{2011MNRAS.415.2101H}, adopting cosmological parameters ($\Omega_m$ = 0.288, $\Omega_\Lambda$ = 0.712, $\Omega_b$ = 0.045, $H_0$ = 69.33 $\mathrm{km}s^{-1}\mathrm{Mpc}^{-1}$, $n_s$ = 0.971, and $\sigma_8$ = 0.830), consistent with the WMAP9 results \citep{2013ApJS..208...19H}. The simulation box, with a comoving volume of $(6\, \rm{Mpc})^3$, is initialized with $128^3$ root cells. High-resolution dark matter particles with a mass of $2448 \msun$ are used for the zoom-in region of $(0.52\, \rm{Mpc})^3$ (comoving), which includes two halos with $2 \times 10^9 \msun < M_{halo} < 5 \times 10^9 \msun$ at $z \sim 10$. 
The halos we select are typical halos at z = 10, located within a filament that is relatively isolated compared to the richer structures in the upper-left region of the bottom panel in Fig.~\ref{fig:simulation_figure},
thereby avoiding extreme environmental influences and allowing the results presented to be representative of halos of similar masses at the relevant redshift.
The Lagrangian volume, or mask, is defined by a scalar quantity that is advected passively with the flow throughout the simulation. The halo virial radius is $R_{vir}=4\, \rm{kpc}$, with the gas density and temperature profiles at $z=10$ shown on the left side of Fig.\ref{fig:simulation_profile}. These profiles provide insight into the thermodynamic state of the halo gas and how it evolves under different physical conditions. Initially, the passive scalar takes a value of 1 inside the mask and 0 outside. Refinement is allowed in regions where the passive scalar exceeds $10^{-3}$, provided that at least one of the following conditions is satisfied: i) the number of dark matter particles within the cell exceeds 8, or ii) the total baryonic mass in the cell exceeds $3060 \msun$. These criteria enable the simulation to achieve a maximum spatial resolution of $\Delta x_{\text{min}} \approx 17.9\,\mathrm{pc}$ (physical) at redshift $z = 10$.

We adopt the equilibrium chemistry model for primordial species (H and He) assuming collisional ionization equilibrium in the presence of a homogeneous UV background. The primordial gas is allowed to cool down to $10^4 \mathrm{K}$ through collisional ionization, excitation, recombination, Bremsstrahlung and Compton cooling. Metal-enriched gas can cool further down using cooling rates tabulated by \cite{1993ApJS...88..253S} above $10^4 \mathrm{K}$ and those from \cite{1972ARA&A..10..375D} below $10^4 \mathrm{K}$. The heating of the gas from a uniform UV background takes place after redshift $z_{reion} = 10$, following \cite{2017ApJ...836..216P}. Motivated by the radiation-hydrodynamic simulation results that the UV background is self-shielded in optically thick regions \citep{2012MNRAS.423..344R}, we assume that UV photo-heating rates are reduced by a factor $\mathrm{exp}(-n_H/n_{shield})$, where $n_{shield} = 0.01 \mathrm{H\ cm}^{-3}$. 

We adopt a star formation number density threshold of $n_0 = 65\,\mathrm{H\ cm}^{-3}$ with a star formation efficiency of $\epsilon_\star = 0.1$, modeled using the Schmidt law $\dot\rho_\star = \epsilon_\star \rho / t_{ff}$, where $\dot\rho_\star$ is the star formation rate (SFR) density, and $t_{ff} = \sqrt{3 \pi / 32 G \rho_{\text{gas}}}$ is the local free-fall time of the gas \citep{2006A&A...445....1R, 2008A&A...477...79D}. 
We assume that 21\% of the stellar mass is returned to the surroundings, with 7.5\% consisting of newly synthesized metals; that is, $\eta_{\rm{SN}} = 0.21$. 
Massive stars are modeled as supernovae (SNe) that explode after 10\,Myr. For kinetic feedback, SNe are implemented as Sedov blasts with an initial ejecta radius of 35\,pc. The initial Sedov blast wave propagates with a velocity given by
\begin{equation}
u_{\rm{Sedov}} = \frac{\sqrt{2}}{5} \left[ f_{\rm{ek}} \eta_{\rm{SN}} \left( \frac{\Delta x}{\delta x} \right)^3 \frac{1}{1 + \eta_{\rm{SN}} + \eta} \right]^{1/2} u_{\text{SN}},
\end{equation}
where $f_{\rm{ek}} = 0.05$, $\Delta x$ is the size of the cell where the explosion occurs, $\delta x$ is the blast radius from the explosion center, and $u_{\text{SN}}$ is the velocity corresponding to the kinetic energy of a single SN explosion \citep{2008A&A...477...79D}. This model is used as our fiducial SN feedback scheme.


We also run a simulation based on the mechanical feedback model from \citet{2014ApJ...788..121K}, where SN energy and momentum are injected according to the evolutionary stage of the Sedov-Taylor blast wave, capturing blastwave momentum and energy in all phases of the evolution.
This model accounts for the continuous distribution of massive star lifetimes, ranging from 3 to 40~Myr \citep{2015MNRAS.451.2900K}. 

Additionally, we consider a simulation employing the delayed cooling model introduced by \citet{2013MNRAS.429.3068T}, where radiative cooling is temporarily disabled in SN-heated gas. The key free parameter in this model is $t_{\rm{delay}}$, which sets the timescale for turbulent energy dissipation. Following previous studies \citep{2013MNRAS.429.3068T,2014MNRAS.444.2837R,2015MNRAS.447.1353M,2016MNRAS.457.1722R}, we adopt a fiducial value of $t_{\rm{delay}} = 20\,\mathrm{Myr}$ in our simulation.

Supermasive black holes (SMBHs) are modelled using a sink particle algorithm \citep{1995MNRAS.277..362B,2004ApJ...611..399K,2014MNRAS.445.4015B,2017MNRAS.469..295B}. The formation of sink particles are identified on the fly using the clump finder $PHEW$ \citep{2015ComAC...2....5B} with an initial seed mass of $10^3 \msun$. We form the BH seed with when all of the following conditions are satisfied: i) the halo mass seed is larger than $4\times 10^4 \msun$ ii) the gas clump mass is larger than $10^4 \msun$ iii) 4-cell ball average density larger than SF threshold iv) peak density is larger than three times SF threshold. SMBH accretion follows the Bondi-Hoyle-Lyttleton \citep{1939PCPS...35..405H,1944MNRAS.104..273B,1952MNRAS.112..195B} accretion formalism.

\begin{equation}
\dot{M}_{Bondi}=\frac{4 \pi \rho_{\infty} G^2 M_{\rm SMBH}^2}{\left(\bar{c}_{s}^2 / \beta_{boost}+\bar{v}_{rel}^2\right)^{3 / 2}},
\end{equation}

\noindent
where, \( G \) is the gravitational constant, and \( \rho_{\infty} = \bar{\rho} / \alpha(x_{\rm sink}) \) represents the density at infinity in the classical Bondi accretion solution, where \( \alpha(x_{\rm sink}) \) is the dimensionless density profile of the Bondi self-similar solution \citep{1992JBAA..102..230S} and $x_{\rm sink} = r_{\rm sink}/r_{\rm Bondi}$ is the dimensionless radius evaluated at the sink sphere radius. The averaged density \( \bar{\rho} \), sound speed \( \bar{c}_{s} \), and relative velocity between the black hole (BH) and the gas \( \bar{v}_{\text{rel}} \) are computed within a sphere of radius \( r_{\rm sink} = 4\Delta x_{min} \), with contributions from each cell weighted by \( w \propto \exp(-r^2 / r_K^2) \) \citep{2004ApJ...611..399K}.
The kernel radius $r_{K}$ is determined based on whether the Bondi radius $r_{\rm Bondi} = G M_{\rm SMBH} / (c^2_{s}/ \beta_{boost} + \bar{v}^2_{rel})$ is resolved or not, where $c_s$ and $v_{rel}$ are the sound speed and relative velocity in the cell containing the sink particle. The radius is defined as follows:
\begin{align}
r_{K} = \begin{cases}
\Delta x_{min} / 4 & \text{if } r_{\rm Bondi} < \Delta x_{min} / 4, \\
r_{\rm Bondi} & \text{if } \Delta x_{min} / 4 \leq r_{\rm Bondi} \leq 2 \Delta x_{min}, \\
2 \Delta x_{min} & \text{if } r_{\rm Bondi} > 2 \Delta x_{min}.
\end{cases}
\end{align}

To account for unresolved dense and cold gas clumps, we increase the accretion rate by reducing the average sound speed $c_s$ in the vicinity of the sink with $\beta_{boost}$ as in \cite{2009MNRAS.398...53B}.

In our simulation, we adopt the prescription $\dot{M}_{\rm acc} = \min(\dot{M}_{\rm BHL}, 3\times\dot{M}_{\rm Edd})$ in the super-Eddington case and $\dot{M}{\rm acc} = \min(\dot{M}_{\rm BHL}, \dot{M}_{\rm Edd})$ in the Eddington-limited case, where $\dot{M}_{\rm Edd}$ denotes the Eddington accretion rate. This definition neglects mass loss due to outflows from the inflowing gas. However, the simulation does account for baryon loading from feedback-driven outflows. A drawback of this approach is that it does not strictly enforce ``local mass conservation''. Nonetheless, we argue that, although the method cannot resolve gas flows within the Bondi radius down to the black hole, it effectively tracks BH mass growth and the associated feedback, which is primarily determined by the BH mass itself. In this sense, within the feedback-regulated regime that we explore here, both the BH mass evolution and its impact on the surrounding medium are reasonably captured.

In RAMSES, AGN feedback is implemented through two distinct modes: a kinetic mode, which operates at low accretion rates and is modeled as a jet-like outflow, and a thermal mode, which becomes active at high but sub-Eddington accretion rates and heats the surrounding gas by releasing thermal energy \citep{2012MNRAS.420.2662D}. The transition between these feedback regimes is determined by the bolometric luminosity-to-Eddington ratio, $\lambda_{Edd} \equiv L/L_{Edd}$.

In the sub-Eddington regime, when $\lambda_{Edd} \leq 0.01$, AGN feedback is in kinetic mode. This mode deposits AGN feedback energy into a bipolar outflow with a jet velocity of $10^4\,\mathrm{km\ s}^{-1}$. The outflow is represented as a cylindrical volume with a cross-sectional radius $\Delta x$ and a height of $2 \Delta x$ following the formulation of \cite{2004MNRAS.348.1105O}. The energy is injected as kinetic energy into all cells within the cylinder, as described in \cite{2016MNRAS.456.3915S}, resulting in a total jet feedback energy of

\begin{equation}
\dot{E}_{jet}=\epsilon_{j} \dot{M}_{acc} c^2,
\end{equation}
where $\epsilon_j$ is the efficiency factor of the kinetic feedback with an open angle of $180^{\circ}$. And the typical value of $\epsilon_j$ is 0.1.

When $\lambda_{Edd} > 0.01$, the AGN enters the so-called “quasar” mode. The luminosity of the AGN is calculated as

\begin{equation}
L_{AGN}=\epsilon_{q} \dot{M}_{acc} \epsilon_{r} c^2,
\end{equation}
with $\epsilon_r = 0.1$ represents the accretion disk radiative efficiency \citep{1973A&A....24..337S} and $\epsilon_q = 0.15$ denotes the hydrodynamic coupling efficiency, which has been calibrated in previous works using RAMSES code \citep{2011MNRAS.414..195T,2012MNRAS.420.2662D,2013MNRAS.434..606G}. Typical values of $\epsilon_q$ rang from 0.05 \citep{2005MNRAS.361..776S,2013MNRAS.431..539W} to 0.15 \citep{2009MNRAS.398...53B,2013MNRAS.434..606G}

Isotropic feedback is an idealization that may overestimate its effects, but our simulations, like most cosmological studies of AGN winds, adopt this approach. By varying feedback strengths, we assess the impact of this potential bias.

We run ten zoom-in cosmological simulations with identical initial conditions but varying feedback mechanisms. In some simulations, the black hole accretion rate is Eddington-limited, while in others, super-Eddington accretion is allowed. For the Eddington-limited simulations, we cap the accretion rate at the Eddington limit, whereas in the super-Eddington case the accretion rate is limited to three times the Eddington limit. Detailed information for each simulation is provided in Table~\ref{tab:simulation_table}. To simplify the description, each simulation is assigned a label based on its characteristics. Simulations with super-Eddington accretion are labeled as $\lambda_{Edd} < 3$, while those with Eddington-limited accretion are labeled as $\lambda_{Edd} < 1$. Simulations without any feedback are labeled as ``No feedback'', while those including only supernova (SN) feedback are labeled as ``Only SN''. The default SN feedback model is kinematic; when the momentum-driven feedback model is used, we label it as ``mom'', and when the delayed cooling feedback model is adopted, we label it as ``delay''. Simulations with both SN feedback and active galactic nucleus (AGN) feedback are labeled according to the specific values of the quasar mode efficiency factor, $\epsilon_{q}$, and the radio mode efficiency factor, $\epsilon_{j}$.

The projected dark matter and gas density of the zoom-in cosmological simulation are shown in Fig.~\ref{fig:simulation_figure}. In the lower panel of Fig.~\ref{fig:simulation_figure}, we display the dark matter distribution of the entire $(6 \, \rm{Mpc})^3$ periodic simulation box at $z=10$. We then plot the region of our zoom-in, high-resolution re-simulation with a volume of $52 \, \mathrm{Kpc}^3$ (physical). When zooming in to a much smaller scale, two halos are clearly visible at the center of our zoom-in simulation, with masses $M_{\mathrm{halo}} = 2 \times 10^9 \, \msun$ and $5 \times 10^9 \, \msun$ respectively, surrounded by smaller sub-halos, as shown in the lower right panel. On this smaller scale, we also present the gas projected density distribution in the upper left panel and the gas temperature in the upper right panel. From the gas density plot, we can see that both halos contain dense gas clumps.


\begin{table*}[t]
	\centering
    \caption{The detail feedback or super-Eddington accretion utilized in different simulations and the mass attained from simulations at $z=10$.}
	\label{tab:simulation_table}
	\begin{tabular}{lcccccccr} 
		\hline
        \multirow{2}{*}{} & \multirow{2}{*}{label}
		  & AGN feedback & $\epsilon_{q}$ & $\epsilon_{j}$ & SN feedback & super-Eddington & $M_{total} (10^9)$ \\
          \cline{3-8}
            & & $M_{dm} (10^9)$ & $M_{\star} (10^8)$ & $M_{gas} (10^8)$ & $M_{hot} (10^8)$ & $M_{cold} (10^8)$ & $M_{\rm sink} (10^3)$ \\
		\hline
        \multirow{2}{*}{1} & \multirow{2}{*}{No feedback, $\lambda_{Edd} < 1$} 
         & $\times$ & - & - & $\times$ & $\times$ & 5.63 \\
        \cline{3-8}
        & & 4.75 & 4.77 & 4.01 & 0.18 & 3.83 & 682.20 \\
        \hline
        
        \multirow{2}{*}{2} & \multirow{2}{*}{ No feedback, $\lambda_{Edd} < 3$}
        & $\times$ & - & - & $\times$ & $\checkmark$ &  5.64 \\
        \cline{3-8}
        & & 4.76 & 1.53 & 3.36 & 0.18 & 3.18 & 179100.40  \\
        \hline
        
        \multirow{2}{*}{3} & \multirow{2}{*}{ only SN, $\lambda_{Edd} < 1$}
        & $\times$ & - & - & $\checkmark$ & $\times$& 5.56 \\
        \cline{3-8}
         & & 4.77 & 1.07 & 5.14 & 0.56 & 4.59 & 330.47\\
         \hline
        
        \multirow{2}{*}{4} & \multirow{2}{*}{ only SN, $\lambda_{Edd} < 3$} 
        & $\times$ & - & - & $\checkmark$ & $\checkmark$ &5.59 \\
        \cline{3-8}
        & & 4.78 & 0.32 & 4.74 & 0.50 & 4.24 & 149703.11  \\
        \hline
        
        \multirow{2}{*}{5} & \multirow{2}{*}{  $\epsilon_{q} = 0.15, \epsilon_{j} = 0.05, \lambda_{Edd} < 1$}
        & $\checkmark$ & 0.15 & 0.05 & $\checkmark$ & $\times$& 5.58 \\
        \cline{3-8}
         & & 4.75 & 1.48 & 4.99 & 0.45 & 4.53 & 363.68 \\
         \hline
        
        \multirow{2}{*}{6} & \multirow{2}{*}{ $\epsilon_{q} = 0.15, \epsilon_{j} = 0.15, \lambda_{Edd} < 3$}
        & $\checkmark$ & 0.05 & 0.05 & $\checkmark$ & $\checkmark$& 5.27 \\
        \cline{3-8}
        & & 4.72 & 0.57 & 4.27 & 0.60 & 3.68 & 709.49  \\
        \hline

        \multirow{2}{*}{7} & \multirow{2}{*}{ $\epsilon_{q} = 0.15, \epsilon_{j} = 0.05, \lambda_{Edd} < 3$}
        & $\checkmark$ & 0.15 & 0.05 & $\checkmark$ & $\checkmark$ & 5.25 \\
        \cline{3-8}
         & & 4.70 & 0.56 & 4.51 & 0.79 & 3.72 & 831.19  \\
         \hline
        
        \multirow{2}{*}{8} & \multirow{2}{*}{ $\epsilon_{q} = 0.05, \epsilon_{j} = 0.05, \lambda_{Edd} < 3$}
        & $\checkmark$ & 0.15 & 0.15 & $\checkmark$ & $\checkmark$ & 5.21 \\
        \cline{3-8}
        & & 4.69 & 0.40 & 4.30 & 0.58 & 3.72 & 347.52  \\
         \hline

         \multirow{2}{*}{9} & \multirow{2}{*}{ $\epsilon_{q} = 0.15, \epsilon_{j} = 0.15, \lambda_{Edd} < 3$ mom}
        & $\checkmark$ & 0.15 & 0.15 & $\checkmark$ & $\checkmark$ & 5.35 \\
        \cline{3-8}
        & & 4.83 & 0.23 & 4.99 & 0.7 & 4.3 & 211.46  \\
         \hline

         \multirow{2}{*}{10} & \multirow{2}{*}{ $\epsilon_{q} = 0.15, \epsilon_{j} = 0.15, \lambda_{Edd} < 3$ delay}
        & $\checkmark$ & 0.15 & 0.15 & $\checkmark$ & $\checkmark$ & 5.36 \\
        \cline{3-8}
        & & 4.85 & 0.12 & 4.94 & 1.27 & 3.67 & 217.67  \\
         \hline
         
	\end{tabular}
\end{table*}

\section{Result} \label{sec:result}

\subsection{Two phase of SMBH growth}
\label{sec:two-phase}


Based on our zoom-in simulations, we present the black hole mass growth histories for all eight runs on the right side of Fig.~\ref{fig:simulation_profile}. These growth curves reveal two distinct phases. Initially, the black hole is seeded with a mass of $10^3\,\msun$ and undergoes rapid growth via super-Eddington accretion, in contrast to the more gradual growth observed in the Eddington-limited cases. We refer to this initial period as the super-Eddington phase.

To examine the effects of AGN feedback during this phase, we analyze results from three simulations with varying quasar-mode coupling efficiencies ($\epsilon_q$) and radio-mode feedback efficiencies ($\epsilon_j$), both within commonly adopted parameter ranges. Despite differences in feedback strength, all simulations undergoing super-Eddington accretion follow a similar trajectory: the black hole grows rapidly, albeit briefly, reaching masses between $10^4$ and $10^5\,\msun$.

This early phase is characterized by efficient black hole growth, sufficient to produce seed masses of $10^4$–$10^5\,\msun$. Such growth appears to be a generic outcome at very high redshifts, as our simulations were not explicitly tuned to promote it. This suggests that early black holes formed via super-Eddington accretion may naturally serve as seeds for the formation of billion-solar-mass SMBHs observed at later cosmic times.

Following this initial phase, the black hole transitions into a self-regulating phase, during which AGN feedback—when super-Eddington accretion is permitted—emerges as the primary mechanism suppressing further growth. This effect is significantly stronger than in simulations with no feedback or only supernova (SN) feedback. As a result, the accretion rate is effectively reduced from super-Eddington to sub-Eddington levels, thereby limiting continued mass accumulation.

This indicates that if the SMBH growth in this second phase is SMBH demand driven, not external gas supplied driven: a faster growth is followed by a longer dormant period, vice versa. Of course, when at some point in time the infalling gas runs out, the SMBH stops growing.

However, the excessive earlier AGN feedback triggered during the super-Eddington phase significantly depletes the cold gas reservoir surrounding the central black hole at later times. As a result, a brief super-Eddington jump-start fails to be retained and
the black hole accreting steadily at the Eddington limit eventually surpasses it by $z \sim 10$. The continuous, Eddington-limited mode thus represents the fastest sustainable pathway to maximal SMBH growth. But given that, in our simulation, black holes follow a two-phase growth trajectory, they can reach $\sim10^6$$\msun$ in $10^{10}$$\msun$ halos by $z = 10$.  Compared to the $\sim10^4\,\msun$ intermediate-mass black holes (IMBHs) commonly invoked for dwarf galaxies \citep{2014AJ....148..136M}, such two phase SMBH growth scenario yields significantly more massive seeds. These could potentially serve as progenitors for Little Red Dots (LRDs), as suggested by recent studies focusing on slightly larger host halos \citep{2024arXiv241214246T,2025MNRAS.536.3677A}.

\subsection{Super-Eddington accretion phase}
\label{sec:super-Eddington}

The most prominent feature of the super-Eddington accretion phase is the rapid growth of the SMBH seed. Regardless of the strength of the AGN feedback parameter within a reasonable range or the specific SN feedback model employed, the black hole growth follows a similar trajectory. By examining the BH accretion rate and its evolution over time, we can gain a clear picture of the accretion history during the super-Eddington phase. On the right side of Fig.~\ref{fig:simulation_profile}, we plot the evolution of the black hole accretion rate, $\dot{M}_{acc}$, as a function of cosmic time for different simulations. We also show the ratio of the BH accretion rate to the Eddington accretion rate, $\lambda_{Edd} = \dot{M}_{acc} / \dot{M}_{Edd}$.

When the ratio $\lambda_{Edd} < 0.01$, the black hole enters the radio mode, where AGN feedback is kinetic and modeled with jet-like outflows. When $\lambda_{Edd} > 0.01$, the AGN enters quasar mode, and the feedback is thermal. During the super-Eddington phase, we observe that the BH accretion rate in all simulations reaches its maximum value for nearly the entire duration of the super-Eddington phase. For the Eddington-limited case, $\lambda_{Edd} \simeq 1$, and for super-Eddington cases, $\lambda_{Edd} \simeq 3$. While the implementation of AGN feedback prevents the black hole from maintaining the maximum accretion rate throughout the super-Eddington phase, it intermittently enters a sub-Eddington phase for short periods. However, simulations with feedback still spend the majority of their time at the maximum accretion rate during the super-Eddington phase.

This rapid growth process is facilitated by the large amount of cold gas present in the center of the halo at early times, which provides sufficient fuel for the black hole to grow at a very high accretion rate.

Accretion is governed by the balance between gas cooling, which promotes infall, and feedback heating, which counteracts it. When the gas cool down, it falls into the central black hole region with free fall time scale $t_{ff}$. While as $M_{\rm sink}$ increase, feedback energy becomes more significant, heating the gas and preventing further infall. This creates a competition between cooling and heating. Initially, when $M_{\rm sink}$ is small, feedback energy is weak, allowing cooling to dominate and enabling efficient black hole accretion. As $M_{\rm sink}$ grows, feedback energy strengthens, eventually surpassing cooling, leading to a decline in the accretion rate. This self-regulation mechanism stabilizes black hole growth.

To physically estimate the black hole mass at this turning point, we consider the energy balance equation within the sink sphere. This involves formulating the energy equation for the average specific internal energy of the gas within the sink sphere as follows:
\begin{equation}
\frac{\rm{d} U}{\rm{d} t}={L_{\rm{AGN}}} - \int \Lambda(T,Z) n^2 dV,
\label{eq:source}
\end{equation}
where $U$ is the total internal energy of the relevant central region,
$\Lambda(T,Z)$ the standard temperature and metallicity dependent cooling rate,
the last term on the right hand side of the equation 
the cooling rate integrated over a relevant volume surrounding the AGN,
which is shown in panel (c) of Figure 3 to be dominated by the inner region, 
indicating that the cooling rate of
the region surrounding the black hole, and 
hence the solution to the equation is convergent and only weakly dependent on the outer regions.
In Eq.\ref{eq:source} we have neglected the pdV work term for simplicity
without a major loss of accuracy since feedback and cooling terms
tend to dominate.

\begin{figure*}[htbp]
    \center
    \includegraphics[width=\textwidth]{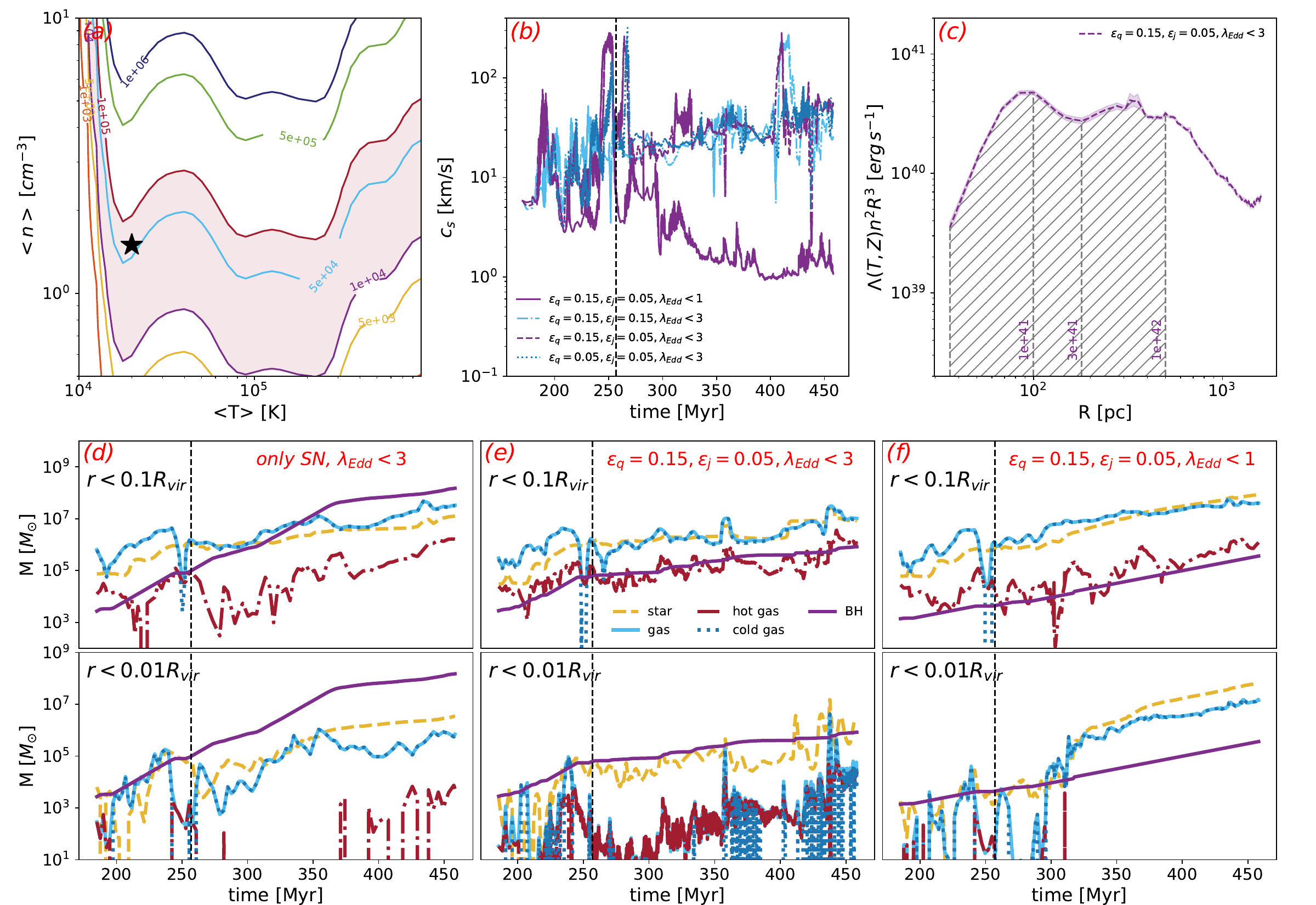} \\
    \caption{(a) The distribution of the transition black hole mass $M_{\rm sink}^t$ determined by the average gas density $<n>$, and the average gas temperature $<T>$. The black star represents the typical value obtained from our simulations, while the red region highlights the range where $M_{\rm sink}^t$ falls between $10^4 - 10^5 \, M_{\odot}$.(b) The time evolution of the sound speed near the black hole, $c_s$, is shown, while (c) presents the squared gas density in spherical shells as a function of radius centered on the black hole times radius cubed, $n^2 R^3$, where the area under the curve is proportional to the cooling rate contribution. (d, e, f) The evolution of stellar mass, total gas mass, hot gas mass, and cold gas mass over time for two regions: the galaxy region (\(0.1 R_{\rm vir}\)) and the black hole region (\(0.01 R_{\rm vir}\)). Panel (d) corresponds to the SN-only case with \(\lambda_{\rm Edd} < 3\), panel (e) to \(\epsilon_q = 0.15\), \(\epsilon_j = 0.05\), and \(\lambda_{\rm Edd} < 3\), and panel (f) to \(\epsilon_q = 0.15\), \(\epsilon_j = 0.05\), and \(\lambda_{\rm Edd} < 1\). Once SMBHs reach masses of \(10^4 - 10^5 \, M_{\odot}\), their growth transitions to a self-regulated phase, driven by AGN feedback that expels gas from the surrounding region. This process ultimately regulates accretion, with higher accretion rates leading to stronger feedback.}
    \label{fig:sinkpro}
\end{figure*}

First, in the super Eddington accretion phase, cooling
dominates over heating. The Bondi accretion rate is so high that we consider the accretion to be maximum set ($3\times \dot{M}_{Edd}$ for super Eddington case, $\dot{M}_{Edd}$ for Eddington limit case). And $\dot{M}_{Edd}$ is 

\begin{equation}
\dot{M}_{\rm{Edd}}=\frac{4 \pi G M_{\rm{sink}} m_{\rm{p}}}{\epsilon_{\rm{r}} \sigma_{\rm{T}} c}=\frac{M_{\rm{sink}}}{t_{\rm{S}}}
\end{equation}
where $t_{\rm{S}}\simeq 45 {\rm Myr}$ is the Salpeter time.
In AGN quasar feedback mode, we have 
\begin{equation}
L_{\mathrm{AGN}}=\epsilon_{\mathrm{q}} \dot{M}_{\mathrm{acc}} \epsilon_{\mathrm{r}} c^2
\end{equation}
It may be seen that for low mass black holes 
gas cooling may dominate over AGN feedback heating.
This cold gas dominated regime remains,
until the SMBH mass grows to the following value


\begin{equation}
M_{\rm sink}^t = \frac{t_{\rm{S}}}{\xi_s\epsilon_{\rm{q}} \epsilon_{\rm{r}} \rm{c}^2} \int \Lambda(T,Z) n^2 dV.
\label{eq:mbh}
\end{equation}
For the super-Eddington case, $\xi_s = 3$. The typical values of the physical quantities in Eq.~\ref{eq:mbh} can be obtained from our simulations. Panel (b) of Fig.~\ref{fig:sinkpro} shows the sound speed $c_s$, while panel (c) shows $\Lambda (T,Z)\,n^2R^3$ as a function of radius centered on the black hole with $1 \sigma$ range in purple in the super-Eddington phase. We see that  $\Lambda (T,Z)\, n^2 R^3$ peaks at $R \sim 480 \, \mathrm{pc}$, which implies that the integral $\int \Lambda (T,Z)\, n^2 \, dV$ will converge at this radius, with finite contribution from larger radii.

By implementing the integral $\int \Lambda (T,Z)\, n^2 \, dV = 10^{42}~\mathrm{erg\,s^{-1}}$ into Eq.~\ref{eq:mbh}, we find that, in our simulation, the black hole mass at the turning point, $M_{\rm{sink}}^{t}$, reaches $5.3 \times 10^4\,\mathrm{M}_{\odot}$—in full agreement with the analytical estimate shown in panel (a) of Fig.~\ref{fig:sinkpro}.

We can also see how the gas density and temperature influence the transition black hole mass, as shown in panel (a) of Fig.~\ref{fig:sinkpro}. The red region indicates where the transition black hole mass ranges from $10^4 - 10^5 \msun$, with the star marking the typical value obtained from the simulation.



Our analytical analysis also explains why the accretion rate in the Eddington-limit case does not slow down at this turning point. In the Eddington-limit case, the physical quantities discussed above are similar to those in the super-Eddington case. However, the black hole mass is not sufficiently large, and as a result, this turning point does not manifest.

Beyond this turning point, strong feedback causes a dramatic drop in the accretion rate. After this time, even with super-Eddington accretion allowed, the accretion rate in simulations with AGN feedback cannot maintain a continuous super-Eddington state. In contrast, simulations with no feedback or only supernova (SN) feedback can sustain super-Eddington accretion for a longer period. 

It is important to emphasize that the first drop in accretion rate is likely due to SN feedback, as this drop also occurs in the SN-only feedback case. However, this drop delays black hole growth without halting super-Eddington accretion, as the accretion rate rebounds to super-Eddington levels after a short period. In contrast, the second drop, caused by AGN feedback, forces the accretion rate to settle at the Eddington limit, regardless of the strength of the AGN feedback parameters or the specific SN feedback model employed.

Over the past few decades, observational evidence has strongly supported the occurrence of super-Eddington accretion. For example, detections of X-ray binaries such as SS 433 \citep{2002PASJ...54..253O} and ultra-luminous X-ray sources, which may host stellar-mass black holes, are considered signatures of accretion rates exceeding the Eddington limit \citep{2006ApJ...649..730W}. In simulations, super-Eddington accretion is often invoked to explain the formation of exceptionally massive black holes observed in the early Universe \citep{2015ApJ...804..148V, 2017MNRAS.471..589P}. Therefore, our simulations offer a natural mechanism for the formation of SMBH seeds with masses ranging from $10^4 - 10^5 \msun$, without relying on any assumptions, within a brief cosmic timescale of approximately 50-100 Myr for typical halos at high redshift, potentially enabling them to serve as progenitors of LRDs.

\subsection{Self regulation of AGN feedback}
\label{sec:self-regulation}

When entering the self-regulation phase, feedback energy exceeds cooling, preventing the black hole from sustaining a very high accretion rate. Consequently, its growth trajectory closely follows the Eddington accretion rate.

To investigate how AGN feedback self-regulates black hole growth, we present the evolution of the total mass in various physical components—stars (yellow dashed line), gas (light blue solid line), hot gas (red dash-dot line), cold gas (blue dotted line), and black holes (purple solid line)—over time in two regions: the galaxy region ($0.1 R_{vir}$) and the black hole region ($0.01 R_{vir}$). This evolution is shown in Fig.~\ref{fig:sinkpro} (d, e, f).


In Fig.~\ref{fig:sinkpro}, we compare the simulation with only supernova (SN) feedback (panel d) to the case that includes both supernova and AGN feedback with super-Eddington accretion (panel e).
When SN feedback is introduced in the super-Eddington accretion case (panel d), we see a decline in cold gas before the transition from the super-Eddington phase to the self-regulation phase. This drop corresponds to the first decline described in Sec.~\ref{sec:super-Eddington}. 

Following this, the cold gas mass begins to increase. Since the heating effect of SN feedback is temporary, the black hole resumes its growth, and this process does not hinder its long-term accretion. However, when AGN feedback is included, as shown in panel (e), a key difference emerges: at the transition between the super-Eddington accretion phase and the self-regulation phase, the re-cooling of cold gas observed in the SN-only case disappears from the black hole region.

As the hot gas component also decreases and the stellar component does not show significant growth, the most plausible explanation for the disappearance of cold gas is that it is being expelled from the black hole region by AGN feedback. This feedback mechanism drives both cold and hot gas outward, preventing cold gas from returning to the black hole region for an extended period. Without a sufficient supply of cold gas as fuel, super-Eddington accretion can no longer be sustained, causing the black hole accretion rate to drop back to the Eddington limit.
By comparing the super-Eddington case (panel e) with the Eddington-limited case that includes AGN feedback (panel f) in Fig.~\ref{fig:sinkpro}, we find that while the black hole initially grows rapidly during the super-Eddington phase, its growth rate declines sharply as AGN feedback expels gas from the surrounding region. This ultimately leads to a final black hole mass comparable to that in the Eddington-limited scenario.

While super-Eddington accretion enables rapid early growth, it also triggers strong AGN feedback, which in turn suppresses further accretion. In other words, the black hole self-regulates its growth through feedback-driven gas depletion. This prolonged gas expulsion is consistently observed across all three super-Eddington simulations, despite variations in the feedback efficiency parameters $\epsilon_q$ and $\epsilon_j$, indicating that the self-regulation phase is a robust and universal feature of super-Eddington growth.

Owing to strong AGN feedback during the self-regulation phase, black hole growth in the super-Eddington case proceeds more slowly than in the Eddington-limited case, consistent with the findings of \cite{2023A&A...670A.180M}. The Eddington-limited black hole surpasses its super-Eddington counterpart in mass by $z=10$ under all feedback strengths we have tested. This suggests that continuous, Eddington-limited accretion represents the fastest sustainable pathway to maximal SMBH growth.

\subsection{Comparison with observations}
\label{sec:observation}

Our simulated halos are of typical size at $z= 10$, only slightly larger than the so-called nonlinear mass predicted by the PS theory at $z=10$ using the equation:
\begin{equation}
\sigma (M_h^{\star}) = \delta_c (z) = 1.686 (1+z)
\end{equation}
and $M_h^{\star} = 2.2 \times 10^9 \, \msun/\mathrm{h}$. The halo in our simulation has a mass of $M_h = 5.6 \times 10^9 \ \msun$, which means that the halos we modeled are commonly found in the Universe. 

\begin{figure*}[htbp]
    \center
    \includegraphics[width=\textwidth]{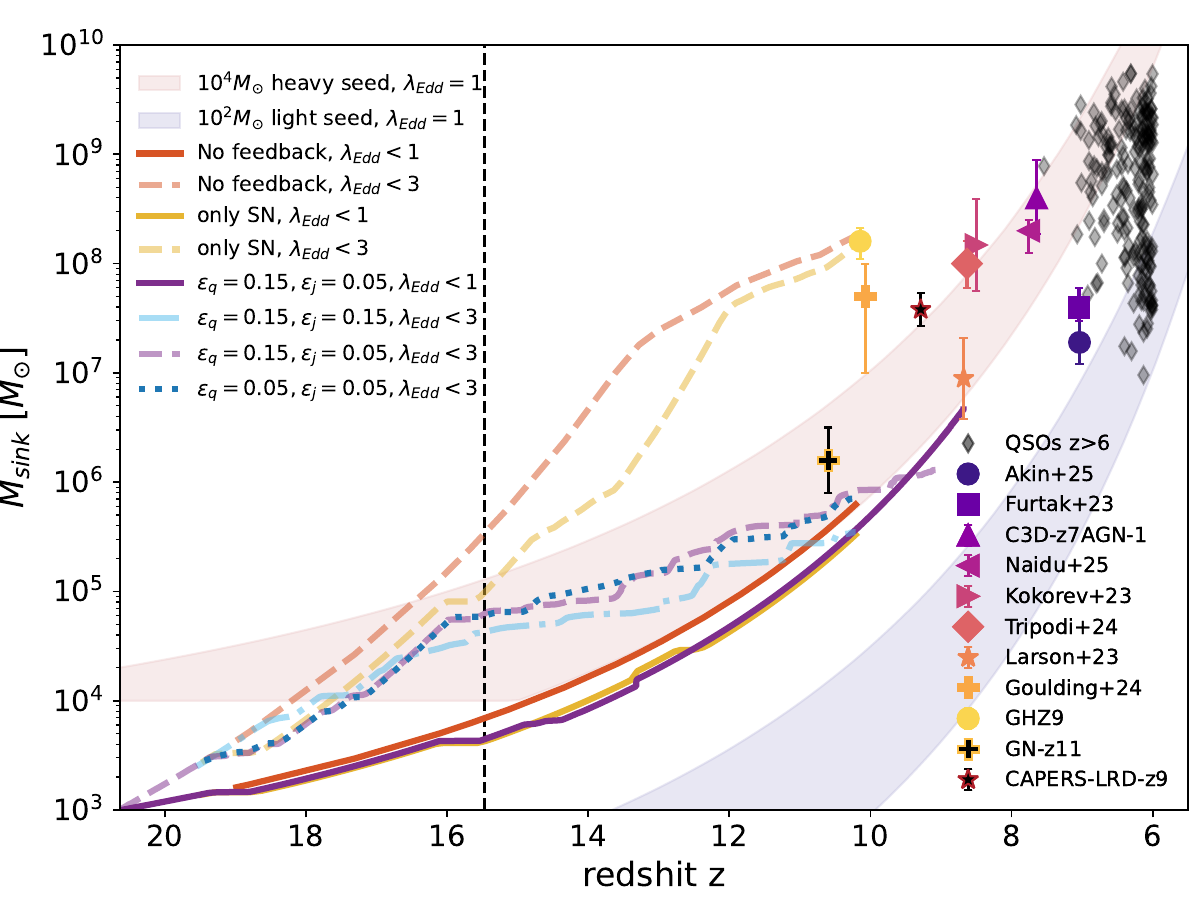}
    \caption{Following \cite{2025arXiv250504609T}, we present the redshifts and black hole masses from our simulations, along with the population of spectroscopically confirmed quasars at $z > 6$. We also include a simple growth model for $10^2\,\msun$ (blue shading) and $10^4\,\msun$ (red shading) black hole seeds accreting at the Eddington limit. A vertical black dotted line marks the transition from the super-Eddington accretion phase to the self-regulation phase. By $z \sim 10$, the SMBH in the more steady, Eddington-limited scenario surpasses its initially super-Eddington jump-start counterpart in mass, indicating that the Eddington-limited mode provides the fastest sustainable pathway for SMBH growth. To account for the $\gtrsim 10^9\,\msun$ quasars observed at $z = 7$–10, seed black holes with initial masses of $10^4$–$10^5\,\msun$ are required.}
    \label{fig:data}
\end{figure*}

Such typical halos provide a valuable framework for understanding the existence of the numerous bright quasars observed at high redshifts. In Fig.~\ref{fig:data}, we present the redshift–black hole mass relation from our simulations, alongside the population of spectroscopically confirmed quasars at $z > 6$. Following a similar approach to \cite{2025arXiv250504609T}, we also include simple analytical growth tracks assuming Eddington-limited accretion (with a 10\% radiative efficiency) for both light seeds ($\sim 10^2\,\msun$) and heavy seeds ($\gtrsim 10^4\,\msun$). The red shaded region illustrates the growth trajectories of heavy seeds forming between redshifts 25 and 15 that begin accreting at the Eddington. For stellar-remnant light seeds, we plot formation redshifts ranging from 30 to 15. The vertical black dotted line denotes the transition from the super-Eddington accretion phase to the self-regulation phase.
Our results indicate that although an initial super-Eddington jump-start enables rapid early growth, the accompanying strong AGN feedback ultimately suppresses further accretion, as detailed in Sec.~\ref{sec:result}. Consequently, the Eddington-limited growth mode emerges as the fastest sustainable pathway for assembling SMBHs by $z = 7$–10.
The most important finding in our analysis is that 
super-Eddington accretion is not the solution to grow 
a stellar mass BH to the observed QSO SMBHs at $z=7-10$.
Seed black holes with initial masses in the range of $10^4$–$10^5\,\msun$ at $z=15-20$ will be needed, with the additional condition that the subsequent growth is at the Eddington rate continuously. 

We emphasize that the actual mechanism of AGN feedback remains uncertain. The model used in our simulations may not fully capture the true nature of feedback. In the limiting case where AGN feedback is negligible or absent, super-Eddington accretion could proceed unimpeded, potentially allowing stellar-mass seeds to grow rapidly into very massive black holes. This could provide a straightforward explanation for the existence of extremely massive black holes, such as the one inferred in GHZ9. But if AGN feedback were entirely ineffective, we would expect an overabundance of overly massive black holes, which is not observed. This implies that strong AGN feedback likely regulates black hole growth in most cases, although an alternative realistic feedback mechanism could potentially replace the one currently assumed, and that the number of AGN feedback-free black holes should remain limited.

Furthermore, black hole mergers can represent an alternative pathway for rapid growth. For instance, incorporating mergers alongside super-Eddington accretion could more easily account for objects like GN-z11. However, this scenario should be considered a supplementary channel and deserves further investigation to evaluate its importance.

\subsection{Comparison with other works}
\label{sec:previous}

This study presents some new findings. 
Here we compare our results with some of previous
works on super-Eddington accretion, where relevant or possible.

\citet{2016MNRAS.459.3738I} perform one-dimensional radiation hydrodynamical simulations to study very high-rate, spherically symmetric accretion flows onto massive black holes. They find that the black hole mass can grow to at least $M_{\rm{BH}} \gtrsim 10^5\,\mathrm{M}_{\odot}$ via hyper-Eddington accretion, independent of the initial seed mass. This argument relies on the assumption that efficient atomic cooling establishes a gas density profile ($\propto r^{-2}$), which is in good agreement with our simulated gas profiles.

Using their analytic expressions:
\begin{equation}
T_{\mathrm{vir}} \simeq 1.9 \times 10^4\, M_{\mathrm{h},8}^{2/3}~\mathrm{K} \left( \frac{1+z}{21} \right),
\end{equation}
and
\begin{equation}
M_{\mathrm{BH}} \lesssim 1.4 \times 10^5\, T_{\infty,4}^{1/2} T_{\mathrm{vir},4}~\mathrm{M}_{\odot},
\end{equation}
and applying our simulation parameters of $T_{\mathrm{vir}} \sim 7 \times 10^4~\mathrm{K}$ and $T_{\infty} \sim 10^4~\mathrm{K}$, we obtain $M_{\rm{BH}} \lesssim 10^6\,\mathrm{M}_{\odot}$. This estimate is slightly higher than our results with $M_{\rm{BH}}\sim 10^5 \msun$, likely due to the inclusion of star formation and SN feedback in our simulations, which can terminate hyper-Eddington accretion earlier and limit black hole growth.

Nonetheless, their analytical treatment of gas profile of halo region on the left side of Fig.\ref{fig:simulation_profile} and the conclusion that the final BH mass is largely independent of the initial seed mass are in good agreement with our simulation results. This suggests that a substantial amount of gas is already present within the halo and available to fuel black hole growth once it reaches the central region.

However, it is important to clarify a key difference in interpretation. While the authors argue that once seed black holes grow to $\gtrsim 10^5\,\mathrm{M}_{\odot}$ via hyper-Eddington accretion, no further rapid growth phase is needed to reach supermassive black hole masses ($\gtrsim 10^9\,\mathrm{M}_{\odot}$) at high redshift, our results indicate otherwise. After the hyper-Eddington phase, strong feedback suppresses further gas inflow, leading to a slower accretion phase. This decelerated growth is eventually overtaken by the Eddington-limited accretion mode case. As a result, we argue that Eddington-limited accretion remains the most efficient long-term growth mode for black hole seeds.

In \citet{2023A&A...670A.180M}, although a different AGN feedback model is adopted—kinetic feedback in the inner region and thermal feedback in the outer region during super-Eddington accretion—and the simulation assumes a massive BH seed of $10^6\,\mathrm{M}_{\odot}$, their results are consistent with ours: super-Eddington accretion does not lead to significantly faster black hole growth compared to the Eddington-limited case. This is because, once the black hole forms as a heavy seed, it already lies beyond the super-Eddington growth phase, and strong feedback mechanisms are already in place, effectively limiting further rapid accretion. 

Our simulations begin with a light BH seed of $10^3\,\mathrm{M}_{\odot}$, enabling us to capture the growth phase and hence the two-phase growth process: an initial super-Eddington accretion phase followed by a transition to sub-Eddington growth. Therefore, the findings in \citet{2023A&A...670A.180M} is consistent with and in some sense  support the conclusion of our two-phase growth scenario where a slower growth phase is inevitable.

\section{Conclusion} \label{sec:conclusion}
In this work, we employ cosmological hydrodynamic simulations to investigate the growth rate of supermassive black holes (SMBHs). We identify a key phenomenon: during the early stages, when super-Eddington accretion is permitted in halos of mass $10^9$–$10^{10}\,\msun$, SMBH growth proceeds through a two-phase process: an initial super-Eddington accretion phase followed by a self-regulation phase.

In the super-Eddington phase, our simulations reveal a new scenario in which SMBH seed masses of $10^4$–$10^5\,\msun$ naturally form within a short cosmic timescale of approximately 50–100 Myr. This early growth is largely insensitive to AGN feedback parameters or the specific SN feedback model employed, highlighting a robust and efficient mechanism for rapid black hole formation at high redshift. The growth rate during this phase significantly exceeds that of the Eddington-limited case. However, once the SMBH reaches a mass of $10^4$–$10^5\,\msun$, it transitions into a self-regulated phase, in which AGN feedback expels cold gas from the vicinity of the black hole and suppresses further accretion to sub-Eddington levels. 

By $z \sim 10$, the SMBH in the Eddington-limited case ultimately overtakes its initially super-Eddington counterpart in mass. This suggests that the continuous, Eddington-limited mode constitutes the fastest sustainable pathway to growing SMBHs to the billion-solar-mass scales observed at $z = 7$–10. To explain the abundance of luminous quasars at these redshifts, seed black holes with initial masses of $10^4$–$10^5\,\msun$ are required. But theses massive seeds have to originated differently than those produced "naturally" during a super-Eddington accretion phase.

The halo modeled in this study represents a typical high-redshift environment, commonly found in the early universe, reinforcing the broader applicability of our findings to the general SMBH population. 

The most important finding in our analysis is that 
super-Eddington accretion is not the solution to grow 
a stellar mass BH to the observed overmassive QSO SMBHs at $z=7-10$.
Seed black holes with initial masses in the range of $10^4$–$10^5\,\msun$ at $z=15-20$ will be needed. 
But theses massive seeds have to originated differently than those that super-Eddington accretion phase appears to "naturally" produce.

\section{Acknowledgments}
\begin{acknowledgments}
We acknowledge the support from the National Key Research and Development Program of China (No.2022YFA1602903), the NSFC \ (No. 11825303, 12347103, 11861131006, 12233005), the science research grants from the China Manned Space project with No. CMS-CSST-2021-A03, CMS-CSST-2021-B01, the Fundamental Research Funds for the Central Universities of China \ (226-2022-00216), the start-up funding of Zhejiang University and Zhejiang provincial top level research support program. R.C. and Z.W. thanks Princeton University for hospitality during a visit when this work was initiated.
We also acknowledge the cosmology simulation database (CSD) in the National Basic Science Data Center (NBSDC) and its funds the NBSDC-DB-10. The simulations and analysis presented in this article were carried out on the SilkRiver Supercomputer of Zhejiang University. 
\end{acknowledgments}

\bibliography{reference}{}
\bibliographystyle{aasjournal}



\end{document}